\documentclass[aps,prb,twocolumn]{revtex4}
\usepackage{epsfig}
\begin{document} 
\title{Quantum and thermal fluctuations in a two-dimensional correlated band ferromagnet 
--- Goldstone-mode preserving investigation with self-energy and vertex corrections} 
\author{Sudhakar Pandey}
\email{spandey@iitk.ac.in} 
\author{Avinash Singh}
\affiliation{Department of Physics, Indian Institute of Technology Kanpur - 208016}

\begin{abstract}
Ferromagnetism in the $t$-$t'$ Hubbard model is investigated on a square lattice.
Correlation effects in the form of self-energy and vertex corrections are systematically incorporated within a spin-rotationally-symmetric scheme which explicitly preserves the Goldstone mode and is therefore in accord with the Mermin-Wagner theorem. Interplay of band dispersion and correlation effects on ferromagnetic-state stability are highlighted with respect to both long- and short-wavelength fluctuations, which are shown to have substantially different behaviour. Our approach provides a novel understanding of the enhancement of ferromagnetism near van Hove filling for $t' \sim 0.5$ in terms of strongly suppressed saddle-point contribution to the destabilizing exchange part of spin stiffness. Finite-temperature electron spin dynamics is investigated directly in terms of spectral-weight transfer across the Fermi energy due to electron-magnon coupling. Relevant in the context of recent magnetization measurements on ultrathin films,
the role of strong thermal spin fluctuations in low dimensions is highlighted, in the anisotropy-stabilized ordered state, by determining the thermal decay of magnetization and $T_c$ within a renormalized spin-fluctuation theory.
\end{abstract}

\pacs{75.50.Pp,75.30.Ds,75.30.Gw}  
\maketitle
\section{INTRODUCTION}

Recent magnetization measurements in ultrathin Fe and FeCo films\cite{Sperl-JAP05, Sperl-JAP06} have highlighted the role of dimensionality and thickness-dependent magnetic anisotropy in controlling thermal excitation of spin waves and thereby determining the temperature fall off of magnetization and hence the transition temperature $T_c$. The temperature dependence of spontaneous magnetization was found to be well described by Bloch's law, and with decreasing film thickness the spin-wave parameter was found to increase significantly compared to the bulk. This stabilization of ferromagnetic (FM) order against thermal fluctuations by magnetic anisotropy in ultrathin transition-metal films, observed even for a single monolayer of Fe on different substrates (Au,W),\cite{Durr-PRL88, Hans-PRB96} is relevant not only from the practical viewpoint of suitability of ferromagnetic materials for magnetic data storage applications,\cite{Heinrich-book} but also in the context of the Mermin-Wagner theorem which implies vanishing $T_c$ for two-dimensional isotropic magnets due to divergent contribution of long-wavelength spin fluctuations and finite but low $T_c$ for small magnetic anisotropy.

With regard to the role of anisotropy on the thermodynamic properties of ultrathin films,
most of the earlier theoretical investigations have been carried out either within the Heisenberg model,\cite{Pescia-PRL90,Mills-PRB91,Hucht-PRB97,Pajda-PRL2000,Pini-PRB05}
where the band-ferromagnetic nature of transition metals is ignored, or within the Hubbard model,\cite{Nolting-PRB98, Nolting-JPCM99} however, without taking into account the collective excitations which are of crucial importance in low-dimensional systems.

Theoretical investigations of collective excitations in the context of transition-metal ultrathin films 
have been carried out extensively in the FM ground state of the Hubbard model including realistic band 
structure.\cite{Mills-JMM98,Mathon-PRB01,Mills-PRB02,Mills-PRB06-03} 
However, transverse spin fluctuations were studied in the random phase approximation (RPA), which neglects correlation effects and overestimates the stability of the ferromagnetic state.

Due to the presence of various competing ground states in the $t$-$t'$ Hubbard model, the issue of ferromagnetic ground state on a square lattice is itself a problem of considerable recent interest. The special favor of the square lattice for the antiferromagnetic ground state at van Hove filling due to the nested Fermi surface is suppressed in the presence of next-nearest-neighbor (NNN) hopping ($t'$), which paves the way for various other instabilities including the ferromagnetic ground state in a narrow density range around the van Hove filling if $t'$ is sufficiently large, as reported recently using a variety of approaches. These include the Hartree-Fock (HF) approximation,\cite{Hirsch-PRB87} quantum Monte Carlo (QMC),\cite{Hirsch-PRB87} the T-matrix approximation,\cite{Hlubina-PRL97,Hlubina-PRB99} a generalized RPA,\cite{Markus-PRB97} the parquet approach,\cite{Irkhin-PRB01} the temperature-cutoff renormalization-group (TCRG),\cite{Honerkamp-PRL01-PRB01} and the two-particle self-consistent (TPSC) approximation.\cite{Hankevych-PRB03} Most of these approaches are limited to the paramagnetic state and (or) at the van-Hove filling. 

It is therefore of interest to theoretically investigate magnetic fluctuations in a low-dimensional band ferromagnet by incorporating correlation effects within an approach in which spin-rotational symmetry and the Goldstone mode are explicitly preserved. The inverse-degeneracy expansion scheme\cite{AS-PRB91} extended recently to study band ferromagnetism in the Hubbard model\cite{AS-PRB06,SP-PRB07} is particularly suitable. In this approach correlation effects are systematically incorporated by including self-energy and vertex corrections so that spin-rotation symmetry and Goldstone mode are preserved order by order. It therefore provides a quantitative description of the long-wavelength, low-energy magnetic fluctuations consistent with the continuous spin-rotational symmetry, which play a dominant role in determining the magnetic behaviour of low-dimensional systems such as ultrathin films and magnetic nanostructures.

Correlation effects incorporated beyond RPA were shown to strongly suppress ferromagnetism, 
as quantitatively demonstrated for several three-dimensional lattices.\cite{SP-PRB07} 
Furthermore, the correlation-induced spin-charge coupling was shown to provide a strong magnon-damping mechanism for modes lying within the Stoner gap,\cite{AS-PRB06} whereas magnon damping within RPA is only due to decay of collective excitations into single particle Stoner excitations.

In this paper we will quantitatively investigate band ferromagnetism within the $t$-$t'$ Hubbard model on a square lattice. We will evaluate quantum corrections to the magnon propagator by incorporating correlation effects in the form of self-energy and vertex corrections within the Goldstone-mode preserving inverse-degeneracy expansion scheme. This will allow us to examine the stability of the ferromagnetic state with respect to both long- and short-wavelength modes. We will show that the enhanced ferromagnetic stability near the van Hove filling for large $t'$ $(\sim 0.5)$ can be readily understood in terms of strongly suppressed saddle-point contribution to the exchange energy term in the spin stiffness. Finally, we will investigate the finite temperature electron spin dynamics directly in terms of the electronic spectral-weight transfer across the Fermi energy, and thereby determine the thermal magnetization decay. In order to highlight the Goldstone-mode preserving character of our approach which is in accord with the Mermin-Wagner theorem, we will focus on the contribution of long-wavelength modes and determine the thermal magnetization decay and $T_c$ in the presence of a small anisotropy gap.

The organization of this paper is as follows. In Sec. II we review the Goldstone-mode preserving diagrammatic approach for quantum corrections to the transverse spin fluctuation propagator. Sections III and IV discuss the stability of ferromagnetic state with respect to long- and short-wavelength fluctuations in terms of the spin stiffness and a characteristic magnon energy for zone-boundary modes, respectively. In Section V we discuss finite-temperature effects in terms of spectral-weight transfer across the Fermi energy due to electron-magnon coupling, and the resulting magnetization decay and $T_c$ in the presence of a small anisotropy gap. Finally, conclusions are presented in Section VI.

\begin{figure}
\begin{center}
\vspace*{-1mm}
\hspace*{-5mm}
\psfig{figure=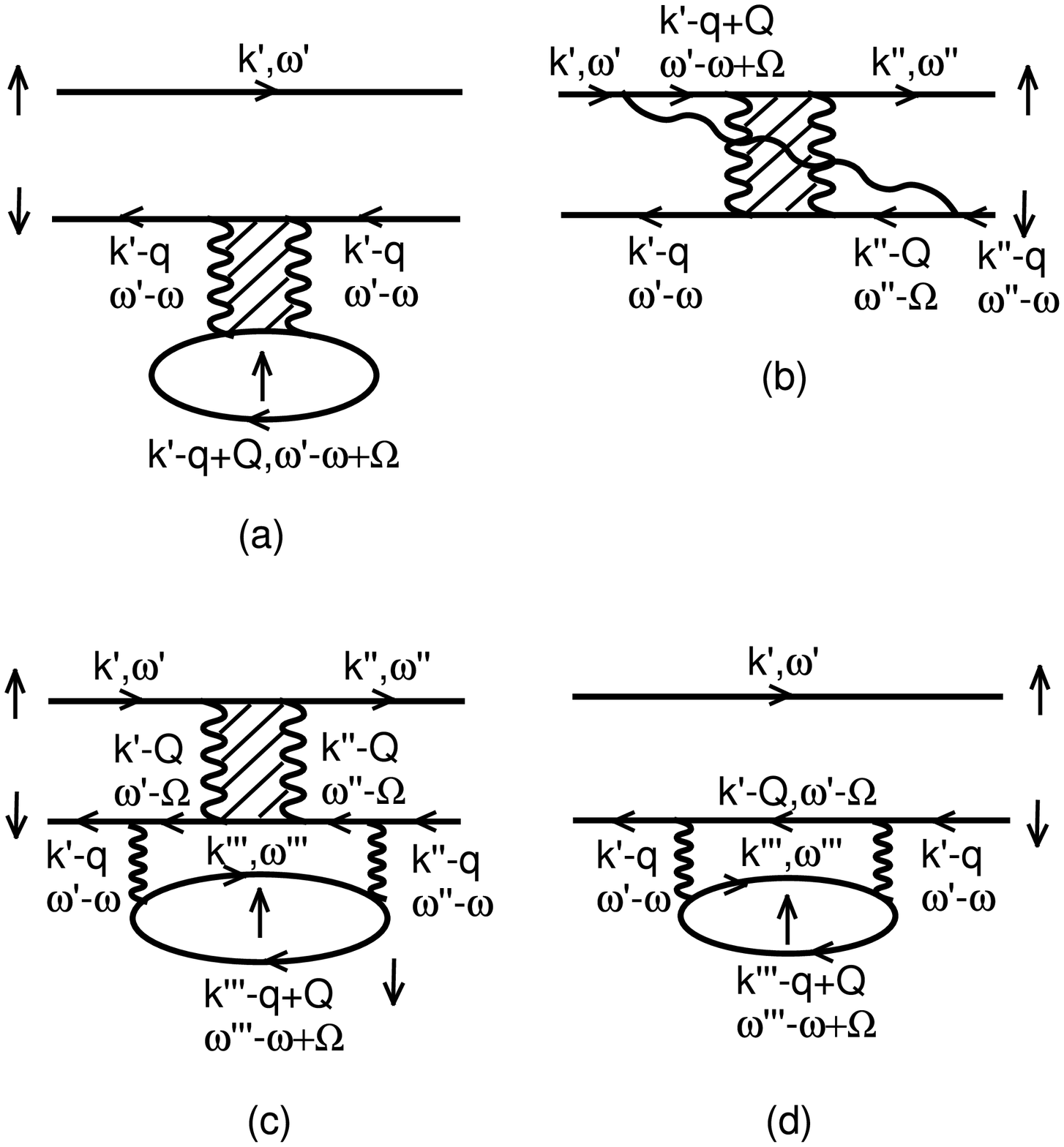,width=90mm}
\vspace*{-15mm}
\end{center}
\caption{First-order quantum corrections to the irreducible particle-hole propagator.}
\end{figure}

\section{Transverse spin fluctuations}

Transverse spin fluctuations are gapless, low-energy excitations in the spontaneously-symmetry-broken state of magnetic systems possessing continuous spin-rotation symmetry, and play an important role in diverse macroscopic properties such as existence of long-range magnetic order, temperature dependence of magnetization, transition temperature, spin correlations etc. We therefore consider the time-ordered, transverse spin-fluctuation propagator 
\begin{equation}
\chi_{ij}^{-+}(t-t') = i \langle \Psi_{\rm G} | T [ S_{i} ^- (t) S_{j} ^+ (t')]|\Psi_{\rm G}\rangle
\end{equation}
in terms of the electron spin-lowering and -raising operators $S_{i} ^\mp = \Psi_{i} ^\dagger (\sigma^\mp/2) \Psi_{i}$,
evaluated in the spontaneously-symmetry-broken ground state $|\Psi_{\rm G}\rangle$. For simplicity, we consider the saturated ferromagnetic state, in which the minority-spin states lie above the Fermi energy; the magnetization $m$ is hence equal to the particle density $n$. 

The spin-fluctuation propagator in ${\bf q},\omega$ space can be expressed generally as 
\begin{equation}
\chi^{-+}({\bf q},\omega) = \frac{\phi({\bf q},\omega)}{1-U\phi({\bf q},\omega)}
\end{equation}
in terms of the exact irreducible particle-hole propagator $\phi({\bf q},\omega)$, which incorporates all self-energy and vertex corrections. The inverse-degeneracy $(1/\cal N)$ expansion\cite{AS-PRB91,AS-PRB06}
\begin{equation}
\phi = \phi^{(0)} + \phi^{(1)} + \phi^{(2)} + ...
\end{equation}
systematizes the diagrams for $\phi$ in powers of the expansion parameter $1/{\cal N}$ which,
in analogy with $1/S$ for quantum spin systems, plays the role of $\hbar$.
Required from spin-rotation symmetry, the exact cancellation $\phi^{(p)}=0$ of all quantum corrections ($p\geq 1$) for $q,\omega=0$ ensures that the Goldstone mode is preserved order by order. 
From (2) it follows that $U^2 \phi^{(p)} ({\bf q},\omega)$ represents the magnon self energy at order $p$.

The bare particle-hole propagator is given by 
\begin{equation}
\phi^{(0)}({\bf q},\omega) \equiv \chi^0 ({\bf q},\omega) =
\sum_{\bf k} \frac{1}{\epsilon_{\bf k-q}^{\downarrow +} - \epsilon_{\bf k}^{\uparrow -} + \omega -i \eta}
\end{equation}
in terms of the HF-level ferromagnetic band energies $\epsilon_{\bf k}^\sigma = \epsilon_{\bf k} - \sigma \Delta$
and the exchange band splitting $2\Delta = mU$; the superscript $+(-)$ refers to particle (hole) states above (below) the Fermi energy $\epsilon_{\rm F}$. As only the "classical" term $\phi^{(0)}$ survives in the $\cal N \rightarrow \infty$ limit, the RPA ladder sum $\chi^0 ({\bf q},\omega)/1-U \chi^0 ({\bf q},\omega)$ amounts to a classical (unrenormalized) description of non-interacting spin-fluctuation modes.

The first-order quantum corrections $\phi^{(1)}$, involving self-energy and vertex corrections of order $1/{\cal N}$,
have been obtained recently for a saturated ferromagnet,\cite{AS-PRB06} and physically incorporate such effects as minority-spin spectral-weight transfer, correlation-induced exchange correction, and coupling of spin and charge fluctuations. Shown diagrammatically in Fig. 1, the quantum corrections were obtained as:

\begin{widetext}
\begin{eqnarray}
\phi^{(1)} ({\bf q},\omega) &=& \phi^{(a)} + \phi^{(b)} + \phi^{(c)} + \phi^{(d)} \nonumber \\
&=&
U^2 \sum_{\bf Q} \int \frac{d\Omega}{2\pi i}
\left [ \left \{ \frac{\chi^0 ({\bf Q},\Omega)}{1-U\chi^0 ({\bf Q},\Omega)} \right \} 
\sum_{\bf k'}\left (\frac{1}{\epsilon_{\bf k' - q}^{\downarrow +} - \epsilon_{\bf k'}^{\uparrow -} + \omega -i \eta} \right )^2 \left ( \frac{1}{\epsilon_{\bf k' - q + Q}^{\uparrow +} - \epsilon_{\bf k'}^{\uparrow -} + \omega - \Omega - i \eta} \right ) \right .
\nonumber \\
&-2 &
\left \{ \frac{1}{1-U\chi^0 ({\bf Q},\Omega)} \right \} \sum_{\bf k'}\left (\frac{1}{\epsilon_{\bf k' - q}^{\downarrow +} - \epsilon_{\bf k'}^{\uparrow -} + \omega -i \eta} \right ) \left (\frac{1}{\epsilon_{\bf k' - q + Q}^{\uparrow +} - \epsilon_{\bf k'}^{\uparrow -} + \omega - \Omega - i \eta} \right )
\nonumber \\
& \times &
\sum_{\bf k''}\left (\frac{1}{\epsilon_{\bf k'' - q}^{\downarrow +} - \epsilon_{\bf k''}^{\uparrow -} + \omega -i \eta}\right ) \left (\frac{1}{\epsilon_{\bf k'' - Q}^{\downarrow +} - \epsilon_{\bf k''}^{\uparrow -} + \Omega -i \eta}\right )
\nonumber \\
&+&
\left \{ \frac{U}{1-U\chi^0 ({\bf Q},\Omega)} \right \} \left \{ \sum_{\bf k'} \left (\frac{1}{\epsilon_{\bf k' - q}^{\downarrow +} - \epsilon_{\bf k'}^{\uparrow -} + \omega -i \eta} \right ) \left (\frac{1}{\epsilon_{\bf k' - Q}^{\downarrow +} - \epsilon_{\bf k'}^{\uparrow -} + \Omega -i \eta}\right ) \right \}^2
\nonumber \\
&\times &
\sum_{\bf k'''} \left (\frac{1}{\epsilon_{\bf k''' - q + Q}^{\uparrow +} - \epsilon_{\bf k'''}^{\uparrow -} + \omega - \Omega - i \eta} \right )
\nonumber \\
&+&
\left . \sum_{\bf k'}\left (\frac{1}{\epsilon_{\bf k' - q}^{\downarrow +} - \epsilon_{\bf k'}^{\uparrow -} + \omega -i \eta}\right )^2 \left (\frac{1}{\epsilon_{\bf k' - Q}^{\downarrow +} - \epsilon_{\bf k'}^{\uparrow -} + \Omega -i \eta}\right ) \sum_{\bf k'''}\left (\frac{1}{\epsilon_{\bf k''' - q + Q}^{\uparrow +} - \epsilon_{\bf k'''}^{\uparrow -} + \omega - \Omega - i \eta} \right ) \right ] \; .
\end{eqnarray}
\end{widetext}

The exact cancellation of $\phi^{(1)}$ for $q,\omega=0$, required from spin-rotation symmetry in order to preserve the Goldstone mode has been discussed earlier.\cite{AS-PRB06} Indeed, the cancellation holds for all $\omega$, indicating no spin-wave amplitude renormalization, as expected for the saturated ferromagnet in which there are no quantum corrections to magnetization.

For finite ${\bf q}$, the magnon energy $\omega_{\bf q}$ is obtained from the pole condition 
$1-U\Re \phi({\bf q},-\omega_{\bf q}) = 0$ in Eq. (2). 
Evaluation of the $\Omega$-integral in (5) was carried out numerically by including the contribution of both the low-energy collective spin-wave excitations as well as the high-energy Stoner excitations, as discussed earlier in detail.\cite{SP-PRB07}

We consider the Hubbard model on a square lattice with band dispersion
\begin{equation}
\epsilon_{\bf k} = -2t (\cos k_x + \cos k_y) + 4t'\cos k_x \cos k_y
\end{equation}
corresponding to NN and NNN hoppings $t$ and $t'$. 
It is instructive to expand the dispersion (6) around special points in ${\bf k}$ space. 
Near $(\pm \pi,0)$, the dispersion exhibits the saddle-point behaviour 
\begin{equation}
\epsilon_{\bf k} = (t+2t')\kappa_y ^2 - (t-2t')\kappa_x ^2 -4t' 
\end{equation}
in terms of the shifted momentum {\boldmath $\kappa$}, with a similar behaviour near $(0,\pm \pi)$ with $\kappa_x$ and $\kappa_y$ interchanged. This saddle-point behaviour results in the characteristic logarithmic density-of-states (DOS) singularity
\begin{equation}
N(\epsilon) \sim \ln \frac{t}{\epsilon + 4t'}
\end{equation}
at energy $-4t'$, as seen in Fig. 2. In the limit $t'\rightarrow t/2$, the effectively one-dimensional dispersion results in a stronger 
($\sim 1/\sqrt{\epsilon+2t}$) DOS singularity at the band bottom. 
Near the origin $(0,0)$ we have 
\begin{equation}
\epsilon_{\bf k} = (t-2t'){\bf k}^2 + t' k_x ^2 k_y ^2 -4(t-t') \; ,
\end{equation}
and the quartic dispersion surviving in the limit $t'\rightarrow t/2$ yields a marginally stronger DOS singularity 
$(1/\sqrt{\epsilon+2t}) \ln \frac{t}{\epsilon+2t}$ at the band bottom. 

\begin{figure}
\begin{center}
\vspace*{-1mm}
\hspace*{-5mm}
\psfig{figure=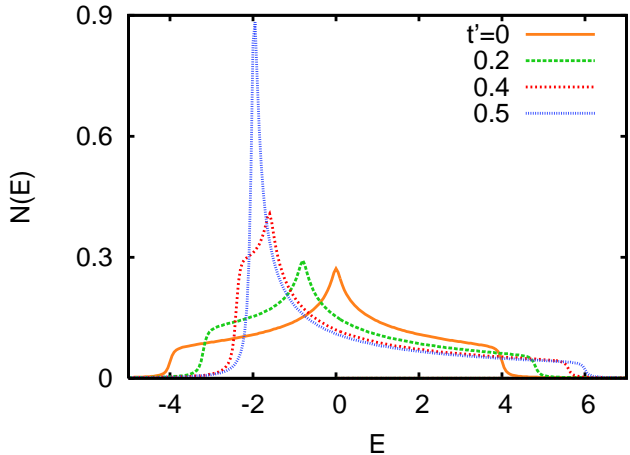,width=90mm}
\vspace*{-1mm}
\end{center}
\caption{DOS for band dispersion (6), showing the enhanced saddle-point contribution and van Hove singularity with increasing next-nearest-neighbor hopping $t'$.}
\end{figure}

We choose an energy scale such that $t=1$. The effects of the distribution of spectral weight in DOS have been investigated for $0 \leq t' \leq 0.5$ in the intermediate range of interaction strengths ($ U/W =$ 0.5, 1.0 and 1.5), 
where the electron bandwidth $W=8t$ is independent of $t'$.  

\section{spin stiffness}

The spin stiffness $D=\omega_{\bf q}/q^2$ in the ferromagnetic state, defined in terms of the magnon energy $\omega_{\bf q}$ for small $\bf q$, provides a quantitative measure of the stability of the ferromagnetic state against long-wavelength fluctuations, with negative $D$ signalling  loss of long-range magnetic order. We first review the different contributions to spin stiffness at the RPA (classical) level as their behavior and interplay provide insight into the magnitude of the first-order quantum corrections and of ferromagnetic-state stability, as discussed in the next subsection.

\subsection{RPA level: hopping and exchange contributions}

Expanding $\chi^0 ({\bf q}, \omega)$ for small ${\bf q}, \omega$ the RPA (classical) spin stiffness can be expressed as\cite{SP-PRB07} 
\begin{equation}
D^{(0)} = \frac{1}{d} \left [\frac{1}{2}
\langle {\mbox{\boldmath $\nabla$}}^2 \epsilon_{\bf k} \rangle  -
\frac{\langle ({\mbox{\boldmath $\nabla$}} \epsilon_{\bf k})^2 \rangle }{2\Delta} \right ]
\end{equation}
in $d$ dimensions.
Here the angular bracket $\langle \; \rangle$ represents momentum summation
normalized over the number of occupied states ($\frac{1}{m} \sum_{\bf k}$).
The two terms in (10) of order $t$ and $t^2/U$
represent hopping and exchange contributions to the spin stiffness, respectively,
corresponding to delocalization-energy loss and exchange-energy gain upon spin twisting.

\begin{figure}
\begin{center}
\vspace*{-17mm}
\hspace*{-3mm}
\psfig{figure=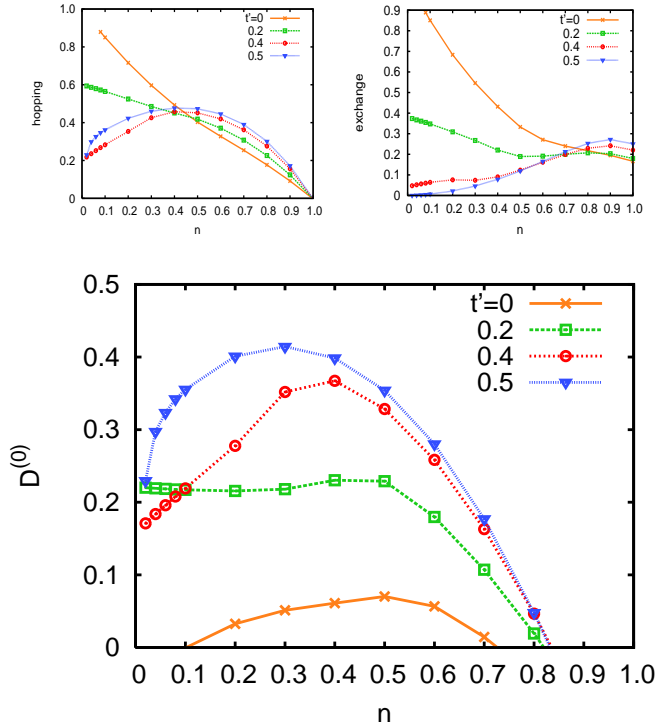,width=90mm}
\vspace*{-20mm}
\end{center}
\caption{Hopping (upper left) and exchange (upper right)
contributions to the RPA spin stiffness (lower) at $U/W=1.5$,
showing enhancement in stiffness and reduction in optimal density with increasing $t'$.}
\end{figure}

The stability of the ferromagnetic state therefore involves a subtle competition between the 
hopping contribution which favors the FM state, and the exchange contribution which tends to destabilize it. Both the hopping and exchange contributions are sensitive to band dispersion and filling, as discussed below. 

Figure 3 shows the hopping and exchange contributions to the spin stiffness. The limit $n\rightarrow 1$ of filled majority-spin band is physically simple and essentially independent of band-dispersion. While the hopping contribution to spin stiffness vanishes, the exchange contribution approaches a finite value, thereby destabilizing the ferromagnetic state.

The low-density ($n \ll 1$) behavior is, however, highly sensitive to the band dispersion and the distribution of spectral weight near the band bottom. With increasing $t'$, although the hopping contribution decreases and nearly saturates, the exchange contribution decreases rapidly and even vanishes as $t'$ approaches 0.5, resulting in substantial enhancement in the stability of the ferromagnetic state. This behaviour of the hopping and exchange contributions can be readily understood from key features of the band dispersion, as explained below.

For low filling, the occupied states lie near ${\bf k}=(0,0)$, so that from Eq. (6)
\begin{eqnarray}
{\mbox{\boldmath $\nabla$}}^2 
\epsilon_{\bf k} &=& 2t (\cos k_x + \cos k_y) -8t' \cos k_x \cos k_y \nonumber \\
&\simeq & (4t-8t') - (t-4t') k^2 \; ,
\end{eqnarray}
leading to a delocalization contribution 
\begin{equation}
D^{(0)} _{\rm deloc} = \frac{1}{4} \langle 
{\mbox{\boldmath $\nabla$}}^2 \epsilon_{\bf k} \rangle = (t-2t') - (t-4t') 2\pi n /4
\end{equation}
to spin stiffness which decreases linearly with filling $n$, with the slope changing sign at $t'= t/4$, as seen in Fig. 3. Here we have used $n=k_{\rm F} ^2 /4\pi$ corresponding to the circular filling pocket. 

Similarly, for the exchange term in the low-density limit, we obtain
\begin{eqnarray}
& & (\mbox{\boldmath $\nabla$} \epsilon_{\bf k})^2 =
(2t-4t'\cos k_y)^2 \sin^2 k_x + x \leftrightarrow y  \\
& \simeq & (2t-4t')^2 [{\bf k}^2 - (k_x ^4 + k_y ^4)/3 ] + 8t' (2t-4t') k_x ^2  k_y ^2
\; , \nonumber
\end{eqnarray}
which yields an exchange contribution 
\begin{equation}
D^{(0)} _{\rm exch} 
= \frac{4\pi (t-2t')^2}{U} \left [ 1 - \frac{2\pi n}{3}
\left ( 1 - \frac{2t'}{t-2t'}\right ) \right ] \; ,
\end{equation}
decreasing {\em quadratically} with $(t-2t')$ and linearly with $n$, 
with the slope changing sign at $t'= t/4$. 

\begin{figure}
\begin{center}
\vspace*{-5mm}
\hspace*{-8mm}
\psfig{figure=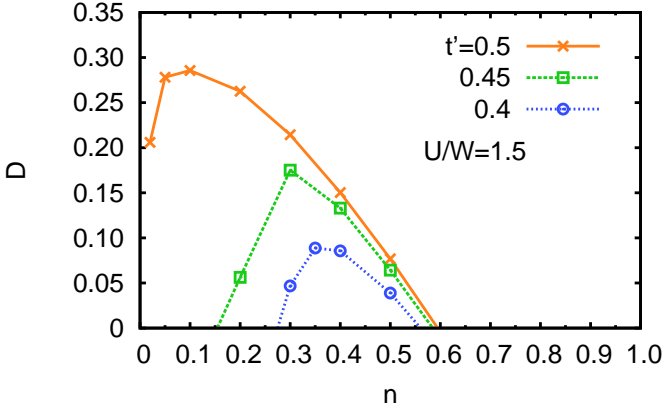,width=90mm}
\vspace*{-1mm}
\end{center}
\caption{The corrected spin stiffness shows a strong enhancement in the low density regime 
as $t'$ approaches 0.5, with optimal density approaching the van Hove filling.}  
\end{figure}

Close to van Hove filling, what is the contribution of states near the DOS singularity to the delocalization and exchange contributions to spin stiffness? As the DOS singularity is associated with saddle points where $\mbox{\boldmath $\nabla$} \epsilon_{\bf k} = 0$, states near the saddle points necessarily yield a small exchange contribution and a finite delocalization contribution. With increasing $t'$, the fraction of states near the saddle points increases, and the exchange contribution correspondingly decreases rapidly, as seen in Fig. 3. Indeed, for $t'=t/2$, the gradient $\mbox{\boldmath $\nabla$} \epsilon_{\bf k}$ vanishes for all states on the two axes $k_x=0$ and $k_y=0$, as seen from (13), and as all these states lie at the band bottom $(-2t)$, the exchange contribution vanishes as $n \rightarrow 0$.    

With increasing $U$, both the density range and stiffness of FM state increase due to the smaller exchange contribution, while the hopping contribution and hence the optimal density remains unaffected. Similar effects of $U$ and $t'$ on ferromagnetic tendency have been found by considering the instability of the paramagnetic state within the generalized RPA.\cite{Markus-PRB97} 

As explicitly shown earlier,\cite{AS-PRB06,SP-PRB07} the first-order quantum corrections to spin stiffness essentially involve correlation-induced exchange processes, and therefore the behavior of classical exchange contribution actually provides a qualitative idea of the first-order quantum corrections as well. The above analysis therefore provides insight into the origin of the enhancement of ferromagnetism near the van Hove filling at sufficiently large $t'$.

\subsection{Quantum correction}
The expansion of the first-order quantum correction $\phi^{(1)}$ for small $q$ has been carried out earlier,\cite{AS-PRB06,SP-PRB07} and yields the corresponding quantum correction to spin stiffness
\begin{widetext}
\begin{eqnarray}
D^{(1)} = 2\Delta U \phi^{(1)} / q^2 
&=& \frac{1}{d} \frac{U^3}{(2\Delta)^3} 
\sum_{\bf Q} \int \frac{d\Omega}{2\pi i} 
\left [ \left \{ \frac{\chi^0 ({\bf Q},\Omega)}{1-U\chi^0 ({\bf Q},\Omega)} \right \} \right .
\sum_{\bf k'} 
\frac{( {\mbox{\boldmath $\nabla$}} \epsilon_{\bf k'} )^2 }
{\epsilon_{\bf k' + Q}^{\uparrow +} - \epsilon_{\bf k'}^{\uparrow -} 
- \Omega - i\eta} \nonumber \\
&-2 &  
\left \{ \frac{1}{1-U\chi^0 ({\bf Q},\Omega)} \right \}
\sum_{\bf k'} 
\frac{{\mbox{\boldmath $\nabla$}} \epsilon_{\bf k'}}
{\epsilon_{\bf k' + Q}^{\uparrow +} - \epsilon_{\bf k'}^{\uparrow -} 
- \Omega- i\eta} .
\sum_{\bf k''} 
\frac{{\mbox{\boldmath $\nabla$}} \epsilon_{\bf k''}}
{\epsilon_{\bf k'' - Q}^{\downarrow +} - \epsilon_{\bf k''}^{\uparrow -} 
+ \Omega- i\eta} 
\nonumber \\
&+& 
\left \{ \frac{U}{1-U\chi^0 ({\bf Q},\Omega)} \right \}
\sum_{\bf k'} 
\frac{1}
{\epsilon_{\bf k' + Q}^{\uparrow +} - \epsilon_{\bf k'}^{\uparrow -} 
- \Omega - i\eta}
\left ( 
\sum_{\bf k''} \frac{{\mbox{\boldmath $\nabla$}} \epsilon_{\bf k''}}
{\epsilon_{\bf k'' - Q}^{\downarrow +} - \epsilon_{\bf k''}^{\uparrow -} 
+ \Omega- i\eta } \right )^2
\nonumber \\
&+& 
\left . \sum_{\bf k'} \frac {1}{\epsilon_{\bf k' + Q}^{\uparrow +} - \epsilon_{\bf k'}^{\uparrow -} - \Omega- i\eta}
\sum_{\bf k''} \frac{({\mbox{\boldmath $\nabla$}} \epsilon_{\bf k''})^2}
{\epsilon_{\bf k'' - Q}^{\downarrow +} - \epsilon_{\bf k''}^{\uparrow -} 
+ \Omega- i\eta} \right ] \; .
\end{eqnarray}
\end{widetext}
All four terms in (15) represent correlation-induced exchange processes 
involving minority-spin intermediate states which are transferred down in energy. 
The spectral-weight transfer is a correlation effect 
corresponding to the possibility of a site being unoccupied by a majority-spin electron. 
Quantum correction to the delocalization contribution was shown to vanish
identically.\cite{AS-PRB06,SP-PRB07} Together with the classical exchange contribution (10), the above correlation-induced exchange contribution yields a further reduction in the renormalized spin stiffness $D=D^{(0)} - D^{(1)}$.

\begin{figure}
\begin{center}
\vspace*{-18mm}
\hspace*{-13mm}
\psfig{figure=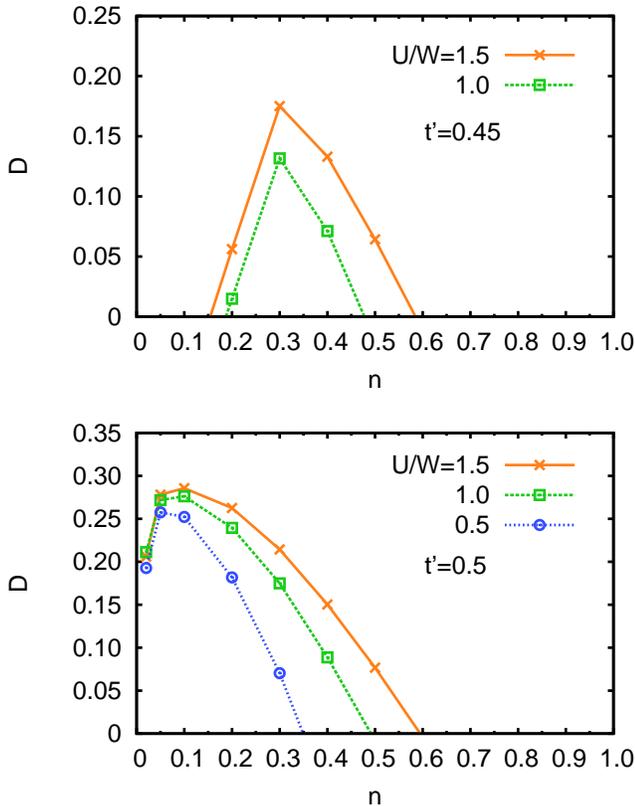,width=100mm}
\vspace*{-15mm}
\end{center}
\caption{The stiffness as well as the stable range of densities are enhanced with increasing $U$ while optimal density remains unaffected. The ferromagnetic state is unstable for $t'\leq 0.35$ at all considered values of $U$, 
for $t'=0.4$ and $U/W \leq 1.0$, and also for $t'=0.45$ at $U/W = 0.5$.}
\end{figure}

Overall, we find that quantum corrections are suppressed by $t'$,
as expected from the decreasing classical exchange contribution (Fig. 3). Nonetheless, quantum corrections are strong enough to destabilize the FM state for $t'\leq 0.35$ for all considered interaction strengths and in the whole range of electronic density. This is in excellent quantitative agreement with results of the T-matrix\cite{Hlubina-PRB99} and generalized-RPA\cite{Markus-PRB97} approaches. 

Figure 4 shows the $D$ vs. $n$ behaviour. 
There is a significant enhancement of ferromagnetism in the low-density regime as $t'$ approaches 0.5. While optimal density decreases and approaches van Hove filling, the stable density range increases due to significant enhancement of spin stiffness in the low density regime. This enhancement of ferromagnetism with increasing $t'$ is in general agreement with results obtained from various other approaches.\cite{Hirsch-PRB87,Hlubina-PRL97,Hlubina-PRB99,Markus-PRB97,Irkhin-PRB01,Honerkamp-PRL01-PRB01,Hankevych-PRB03}       

Figure 5 highlights the interplay of $t'$ and U. 
With increasing $t'$, the ferromagnetic state becomes stable even at relatively small interaction strength. The FM state is unstable in the whole density range for $t'=0.4$ at $U/W \leq 1.0$ and similarly for $t'=0.45$ at $U/W = 0.5$, but becomes stable in a limited density range at intermediate coupling ($U/W$=1.0 and 1.5), with an optimization near $n=0.3$. For $t'=0.5$, quantum correction becomes less significant, resulting in stabilization of the ferromagnetic state even at $U/W=0.5$, and significant enhancement in spin stiffness with decreasing $n$.

\begin{figure}
\begin{center}
\vspace*{-18mm}
\hspace*{-12mm}
\psfig{figure=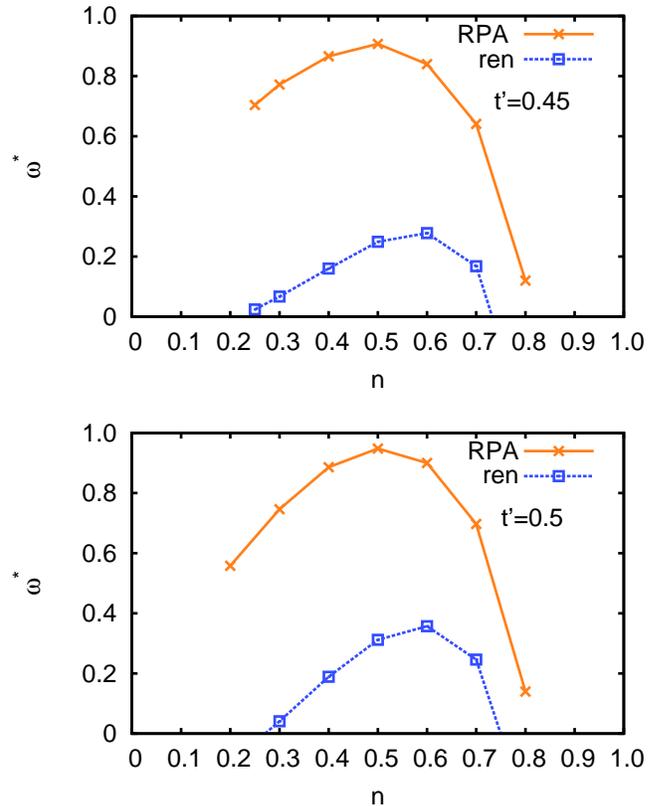,width=100mm}
\vspace*{-15mm}
\end{center}
\caption{Characterizing large-momentum, zone-boundary modes, the renormalized (ren) dominant-mode energy shows a strong suppression at low densities due to quantum corrections, restricting stability to an intermediate range of densities, shown for $U/W=1.0$.}
\end{figure}

For both $t'=0.45$ and 0.5, increasing $U$ results in enhancement of spin stiffness as well as of the stable range of densities. This is because of both enhanced RPA stiffness due to reduced exchange contribution and also reduced quantum corrections.

Our results for the stability of the ferromagnetic state for $t'=0.4$, $0.45$, and $0.5$ are in good quantitative agreement with the prediction of the T-matrix approximation\cite{Hlubina-PRB99} in the low-density regime which, however, predicts a relatively large upper density limit. This quantitative difference with respect to the upper density limit may be attributed to the over-estimation by the T-matrix approach which is more reliable in the low density limit.\cite{Hlubina-PRL97,Kanamori-PTP63,Mattis-PL85} Furthermore, our results for $t'=0.5$ are in excellent quantitative agreement with the predictions of the generalized RPA.\cite{Markus-PRB97} 
Our overall finding of decreasing critical value of $t'$ with increasing $U$ beyond which the ferromagnetic state is stable is in good qualitative agreement with the approaches mentioned in the Introduction. 

\section{Magnon-energy renormalization for large-momentum modes}

We now consider quantum corrections for zone-boundary modes with energy $\omega_{\bf q}$ near the magnon DOS peak energy ($\omega^*$). The behavior of this "dominant-mode" energy $\omega^*$ with electron filling $n$ is of interest as it determines the stability of the FM state with respect to typical short-wavelength fluctuations. We have checked for several such ${\bf q}$ and find the behavior to be quantitatively very similar. 

Figure 6 shows the behavior of the renormalized dominant-mode energy with electron filling for $t'=0.45$ and 0.5, along with the RPA results. In contrast to Fig. 5, the nearly similar behavior shows the relatively much weaker sensitivity of large-$\bf q$ modes on details of band dispersion. The strong suppression at low densities, a trend already present at the RPA level, results in an intermediate-density range of stability with respect to short-wavelength fluctuations.

Comparison of Figs. 5 and 6 shows a marked contrast in the behavior of magnon energy for long- and short-wavelength modes, especially for $t'=0.5$. While positive spin stiffness in the low-density range indicates stability with respect to long-wavelength fluctuations, the vanishing of dominant-mode energy $\omega^*$ for relatively short-wavelength modes
signals the spontaneous onset of strong local fluctuations. On the other hand, while $\omega^*$ shows a peak near  $n \approx 0.6$, indicating stability with respect to local fluctuations, the vanishing spin stiffness near $n = 0.6$
indicates loss of long-range magnetic order. That the stability of the ferromagnetic state with respect to 
long- and short-wavelength fluctuations can differ so substantially is a significant feature of our approach. 

The substantial decrease in magnon energy at large momenta in the low-density regime (Fig. 6) is of considerable relevance in the context of experimentally observed "anomalous softening" in ferromagnetic manganites.\cite{Ye-PRL06}

In the intermediate-density range where FM state is locally stable, but unstable with respect to long-wavelength fluctuations, some commensurate spin density wave states other than ferromagnetic and antiferromagnetic were found in 
the generalized RPA\cite{Markus-PRB97} investigation.   

\section{Spectral-weight transfer and thermal decay of magnetization}
As spin-rotation symmetry and the Goldstone mode are explicitly preserved in our approach, it is therefore in accord with the Mermin-Wagner theorem. Long wavelength modes in one and two dimensions yield divergent thermal fluctuations, indicating absence of long-range magnetic ordering at any finite temperature. With quantum corrections included to yield the correct magnon energy scale for long wavelength modes, our approach therefore provides a quantitative tool for investigating fluctuation effects in low-dimensional band ferromagnetic structures. As a simple illustration, we consider in this section the reduction in magnetization with temperature due to thermal excitation of magnons, suppression of the infrared divergence by anisotropy induced gap in the magnon spectrum, and the consequent stabilization of the ferromagnetic state in two dimensions. 

\begin{figure}
\begin{center}
\vspace*{-0mm}
\hspace*{-18mm}
\psfig{figure=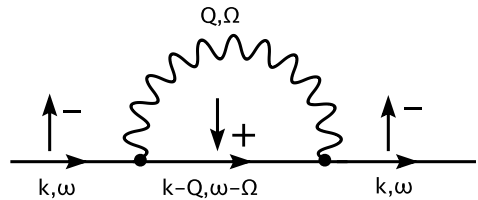,width=70mm}
\vspace*{-45mm}
\end{center}
\caption{The first-order correction to spin-$\uparrow$ electron Green's function due to electron-magnon interaction, which results in spectral-weight transfer across the Fermi energy and thermal magnetization decay.}
\end{figure}

We first consider a simple description of the magnon propagator at finite temperature. Including both quantum and thermal corrections, the magnon propagator is approximately given by
\begin{equation}
\chi^{-+}({\bf Q},\Omega) = \frac {\tilde{m}_{\bf Q}}
{\Omega + \tilde{\Omega}_{\bf Q} - i\eta} \; ,
\end{equation}
where 
\begin{eqnarray}
\tilde{m}_{\bf Q} &=& m_{\bf Q}.\langle S_z \rangle_T / \langle S_z \rangle_0 \nonumber \\
\tilde{\Omega}_{\bf Q} &=& \Omega_{\bf Q}.\langle S_z \rangle_T / \langle S_z \rangle_0
\end{eqnarray} 
are the renormalized magnon amplitude and energy, 
with $\langle  \rangle_T $ denoting the thermal average at temperature $T$. 
These thermal renormalizations arise from the O($\omega$) term in the quantum corrections, which is proportional to the magnetization  $\langle n_\uparrow \rangle_T - \langle n_\downarrow \rangle_T = 2\langle S_z \rangle_T$. At zero temperature, quantum corrections to the O($\omega$) term cancel identically, as expected from the absence of any quantum correction to magnetization. The above form of the magnon propagator, similar to that for a quantum Heisenberg ferromagnet, ensures that the sum-rule 
\begin{equation}
\langle [S^+ , S^-] \rangle_T = 2\langle S_z \rangle_T 
\end{equation}
following from the spin commutation property is identically satisfied at all temperatures. 
The gapped Stoner excitations, which distinguish a band ferromagnet, have been neglected in the above consideration as our focus is on long-wavelength, low-energy fluctuations.

The thermal reduction of magnetization in a band ferromagnet arises from the electron spectral-weight transfer across the Fermi energy. Due to the spin-flip scattering of electrons accompanying the thermal magnon excitations, a portion of the spin-$\uparrow$ spectral weight is transferred to the spin-$\downarrow$ band above the Fermi energy, while an equal amount of spin-$\downarrow$ spectral weight is transferred to the spin-$\uparrow$ band below the Fermi energy. The corresponding changes in the electron densities due to the first order fluctuation process (Fig. 7) are obtained as: 
\begin{equation}
\delta n_\downarrow = -\delta n_\uparrow =
- \sum_{\bf k} \int_{\epsilon_{\rm F}} ^{\infty} \frac{d\omega}{\pi} {\rm Im} 
[G_0 ^\uparrow ({\bf k},\omega) \Sigma^\uparrow ({\bf k},\omega) G_0 ^\uparrow ({\bf k},\omega)] \; ,
\end{equation}
where $G_0 ^\uparrow$ is the HF-level (advanced) propagator and the electron self-energy correction due to electron-magnon coupling is given by 
\begin{widetext}
\begin{eqnarray}
\Sigma^\uparrow ({\bf k},\omega) 
&=& U^2 \sum_{\bf Q} \int \frac {d\Omega}{2\pi i} 
\chi^{-+}({\bf Q},\Omega) 
G_0 ^\downarrow ({\bf k-Q},\omega-\Omega)
\left ( \frac{-1}{1-e^{-\beta\Omega} }\right )
\nonumber \\
&=& U^2 \sum_{\bf Q} \frac{\tilde{m}_{\bf Q}}{e^{\beta \tilde{\Omega}_{\bf Q}} -1} 
\left ( \frac{1}{\omega+\tilde{\Omega}_{\bf Q} 
- \epsilon_{\bf k-Q}^{\downarrow +} +i\eta} \right ) .
\end{eqnarray}
\end{widetext}

Evaluating the spectral weight transfer from Eq. (19), we obtain the thermal reduction in magnetization 
\begin{eqnarray}
& & \langle S_z \rangle_0 - \langle S_z \rangle_T = (\delta n_\downarrow -\delta n_\uparrow)/2 = -\delta n_\uparrow \nonumber \\
&=& U^2 \sum_{\bf Q} \frac{\tilde{m}_{\bf Q}} {e^{\beta \tilde{\Omega}_{\bf Q}} -1} 
\sum_{\bf k} \left ( \frac{1} 
{\epsilon_{\bf k-Q}^{\downarrow +} - \epsilon_{\bf k}^{\uparrow -} - \tilde{\Omega}_{\bf Q}} \right )^2 \; .
\end{eqnarray}
Coupled with Eq. (17) for magnon amplitude and energy renormalization, the above equation self-consistently determines the magnetization $\langle S_z \rangle_T$ at temperature $T$. The $Q^2$ dependence of the magnon energy naturally yields divergent reduction in one and two dimensions, and the low-temperature Bloch-law ($T^{3/2}$) fall off in three dimensions. 

As the temperature approaches $T_c$ from below, both the magnetization $\langle S_z \rangle_T$ and the renormalized magnon energy $\tilde{\Omega}_{\bf Q}$ become vanishingly small, and from Eq. (21) we therefore obtain
\begin{eqnarray}
\frac{1}{k_B T_c} 
&=& U^2 \sum_{\bf Q} \frac {m_{\bf Q}}{\Omega_{\bf Q}} \sum_{\bf k} 
\left ( \frac{1}{\epsilon_{\bf k-Q}^{\downarrow +} - \epsilon_{\bf k}^{\uparrow -}} \right)^2 / \langle S_z \rangle_0
\nonumber \\
&\approx & \frac{2}{n} \sum_{\bf Q} \frac {1}{\Omega_{\bf Q}} \; ,
\end{eqnarray}
where we have taken $m_{\bf Q} \approx 2\langle S_z \rangle_0$ and 
$\epsilon_{\bf k-Q}^{\downarrow +} - \epsilon_{\bf k}^{\uparrow -} \approx 2\Delta = nU$ 
for the dominant small-$Q$ modes. 
With the formal identification $n=2S$, this expression is similar to that obtained by mapping the itinerant electron system to an effective Heisenberg model.\cite{Pajda-PRL2000,Nolting-Soludi01} 
However, to our knowledge, this is the first such result for an itinerant ferromagnet which has been obtained directly in terms of the spectral weight transfer.

For a two-dimensional isotropic system with $\Omega_{\bf Q} \sim Q^2$, 
the transition temperature vanishes due to the logarithmically divergent contribution of long-wavelength magnon modes. Presence of an anisotropy gap in the long-wavelength magnon spectrum $\Omega_Q = DQ^2 + \Delta_a$ due to intrinsic spin anisotropy stabilizes the ferromagnetic ordering at finite temperature by suppressing the divergent contribution of long-wavelength modes. Recent magnetization measurements \cite {Sperl-JAP05}  on ultrathin Fe films have indeed found substantial enhancement in thermal excitation of magnons with decreasing anisotropy gap which is sensitive to the interface between film and substrate. With an anisotropy gap, the expression for transition temperature reduces to  
\begin{equation}
T_c \sim nD / \ln \left (\frac{D \Lambda^2}{\Delta_a} \right )\; ,
\end{equation}
where $\Lambda \sim 1$ is an upper-momentum cutoff. This result for anisotropy-driven stabilization of ferromagnetic ordering at finite temperature is in close analogy with similar investigations for localized spin systems.\cite{ Mills-PRB91,Pajda-PRL2000} With quantum corrections included in the spin stiffness $D$, the above result highlights the role of correlation effects on {\em long-wavelength} fluctuations in essentially determining the transition temperature. 

\section{Conclusions} 
In conclusion, we have investigated quantum corrections to collective magnon excitations in a two-dimensional correlated band ferromagnet. Self-energy and vertex corrections were incorporated within a spin-rotationally-symmetric scheme in which the Goldstone mode is explicitly preserved. By studying the ferromagnetic state spin stiffness, we showed that ferromagnetism is strongly enhanced near the van Hove filling for $t'\sim 0.5$, in agreement with earlier studies from the paramagnetic side. This enhancement in ferromagnetism was shown to be a consequence of strongly suppressed saddle-point contribution near the van Hove singularity to the competing exchange part of the spin stiffness.
  
Due to the strong spin-charge coupling, resulting in an anomalous momentum dependence of the magnon self energy, magnon energies for zone-boundary modes were found to be strongly suppressed in the low-density regime. Also obtained earlier in our three-dimensional calculations,\cite{SP-PRB07} this magnon-energy suppression for zone-boundary modes is relevant in the context of experimentally observed "anomalous softening" in ferromagnetic manganites.\cite{Ye-PRL06}

We also studied the finite-temperature electron spin dynamics due to coupling with the fluctuating transverse field of collective excitations in the presence of a small anisotropy gap. The magnetization decay and $T_c$ were obtained directly in terms of the electronic spectral weight transfer, rather than first mapping the itinerant ferromagnet to an effective Heisenberg spin model. The renormalized spin stiffness was shown to essentially determine the energy scale for magnetization decay and $T_c$. The vanishing of $T_c$ in the isotropic limit due to divergent long wavelength fluctuations is in accord with the Mermin-Wagner theorem. The Goldstone-mode preserving approach thus naturally allows for long-wavelength, low-temperature, and low-dimensional studies in the broken-symmetry state.

These results are relevant in the context of recent magnetization measurements on ultrathin transition-metal films, although quantitative comparison with experiments would necessitate incorporation of several realistic 
features such as multiple $3d$ orbitals as well as presence of multilayers and non-magnetic substrates.

\section{Acknowledgments}
One of us (SP) gratefully acknowledges financial support from CSIR.

\end{document}